\documentclass[aps,a4paper,10pt,twocolumn,nofootinbib]{revtex4}
\usepackage[T1]{fontenc}
\usepackage{mathptmx}
\usepackage{amsfonts}
\usepackage{amsfonts}
\usepackage{eufrak}
\usepackage{mathrsfs}
\usepackage[mathscr]{euscript}
\usepackage[dvips]{graphicx}
\usepackage{fancyhdr}
\usepackage{colordvi}

\newcommand{\ImagPart}{\mathbb{I}\mathrm{m}}
\newcommand{\ArgPart}{\mathbb{A}\mathrm{rg}}

\newcommand{\PhiEps}{\phi_\epsilon}
\newcommand{\PhiMu}{\phi_\mu}

\newcommand{\XARXIV}[1]{\href{http://arxiv.org/abs/#1}{arXiv:#1}}

\begin{document}
\title{Negative refraction in active and passive media: a discussion of various criteria}
\author{P. Kinsler}
\email{Dr.Paul.Kinsler@physics.org}
\author{M.W. McCall}
\email{m.mccall@imperial.ac.uk}
\affiliation{
  Blackett Laboratory, Imperial College London,
  Prince Consort Road,
  London SW7 2AZ, 
  United Kingdom.
}

\lhead{
MCOND}
\chead{Negative refraction in active and passive media: a discussion ...}
\rhead{
\href{mailto:Dr.Paul.Kinsler@physics.org}{Dr.Paul.Kinsler@physics.org}
}

\begin{abstract}

There are three widely-used conditions for characterizing 
 negative refraction in isotropic dielectric-magnetic materials.
Here we demonstrate that whilst all the different conditions
 are equivalent for purely passive media, 
 they are distinct if active media are considered.
Further, 
 these criteria can also be applied to negative refraction
 in acoustic materials, 
 where we might replace the dielectric permittivity $\epsilon$ 
 with a bulk modulus $\kappa$, 
 and the magnetic permeability  $\mu$ with the mass density $\rho$.

\end{abstract}

\date{\today}
\maketitle
\thispagestyle{fancy}

Note: 
 This is a reworded \& 
 expanded revision of \cite{Kinsler-M-2008motl}.

%
\section{Introduction}\label{S-intro}

In 1968, 
 Veselago\cite{Veselago-1968spu} investigated
 lossless media with simultaneously negative permittivity
 and permeability; 
 these are commonly referred to as
 negative refractive index (NRI) media.
In NRI media, 
 the phase velocity of electromagnetic-wave propagation is
 anti-parallel to the energy flow, 
 and the phrase Negative Phase Velocity (NPV) propagation is
 sometimes used as an alternative to negative refraction (or NRI)
 \cite{McCall-LW-2002ejp}.
Note also that some alternative definitions of NRI rely on 
 the opposition of phase and group velocities; 
 hence the use in \cite{Kinsler-M-2008-mcrit} 
 of the acronyms NPVE (NPV w.r.t. energy velocity) 
 and NPVG  (NPV w.r.t. group velocity); 
 here we use just NPV, 
 meaning specifically NPVE.
Such materials have even been experimentally realized \cite{Shelby-SS-2001s}. 
As might be expected in a rapidly evolving field of research, 
 a variety of conditions 
 \cite{McCall-LW-2002ejp,Ruppin-via-LCW,Depine-L-2004motl}
for negative refraction have been 
 proposed in the literature.
As summarized in \cite{Depine-L-2004motl}, 
 it was also shown that all the different conditions are equivalent
 in passive media, 
 despite their rather different mathematical forms. 

In this paper we show that the different conditions
 are \emph{not} equivalent when they
 are extended to active media. 
Active media that support NPV propagation must 
 (of course) be dispersive, 
 and, 
 on account of the Kramers-Kronig relations,
 dispersion implies loss.
Therefore, 
 by embracing active media,
 the limitations imposed by loss might be overcome; 
 although Stockman \cite{Stockman-2007prl} has claimed 
 that there are some important restrictions -- 
 albeit restrictions valid only in the case of perfect transparency 
 at the observation frequency\footnote{This constraint, 
 and its applicability, were recently relaxed in \cite{Kinsler-M-2008-mcrit}}.

Although the discussion below is applied explicitly only
 to electromagnetic materials, 
 and considers the material parameters
 to be permittivity $\epsilon$ and permeability $\mu$; 
 we might also apply it to acoustic materials --
 notably a pressure acoustics model 
 (see e.g. \cite{Kinsler-M-2014pra})
 by substituting $\epsilon$, $\mu$
 with bulk modulus $\kappa$ and mass density $\rho$

In isotropic media where magneto-electric effects are absent,
 gain arises from the response to the electric or magnetic fields:
 i.e. from the permittivity, $\epsilon = \epsilon_r + \imath \epsilon_i$, 
 or the permeability, $\mu = \mu_r + \imath \mu_i$, 
 respectively. 
Situations where both, 
 or either of these occur must be distinguished. 
In this paper,
 if both $\epsilon$ and $\mu$ are active, 
 we denote that a \emph{doubly active} medium.
Next, 
 if neither $\epsilon$ or $\mu$ have negative imaginary parts
 (i.e. both are passive), 
 we denote that a \emph{doubly passive} (or just passive) medium.
If $\epsilon$ is active but not $\mu$, 
 or not $\epsilon$ passive but $\mu$ active,
 we denote that a \emph{singly active} medium.

\begin{figure}
\includegraphics[width=0.75\columnwidth]{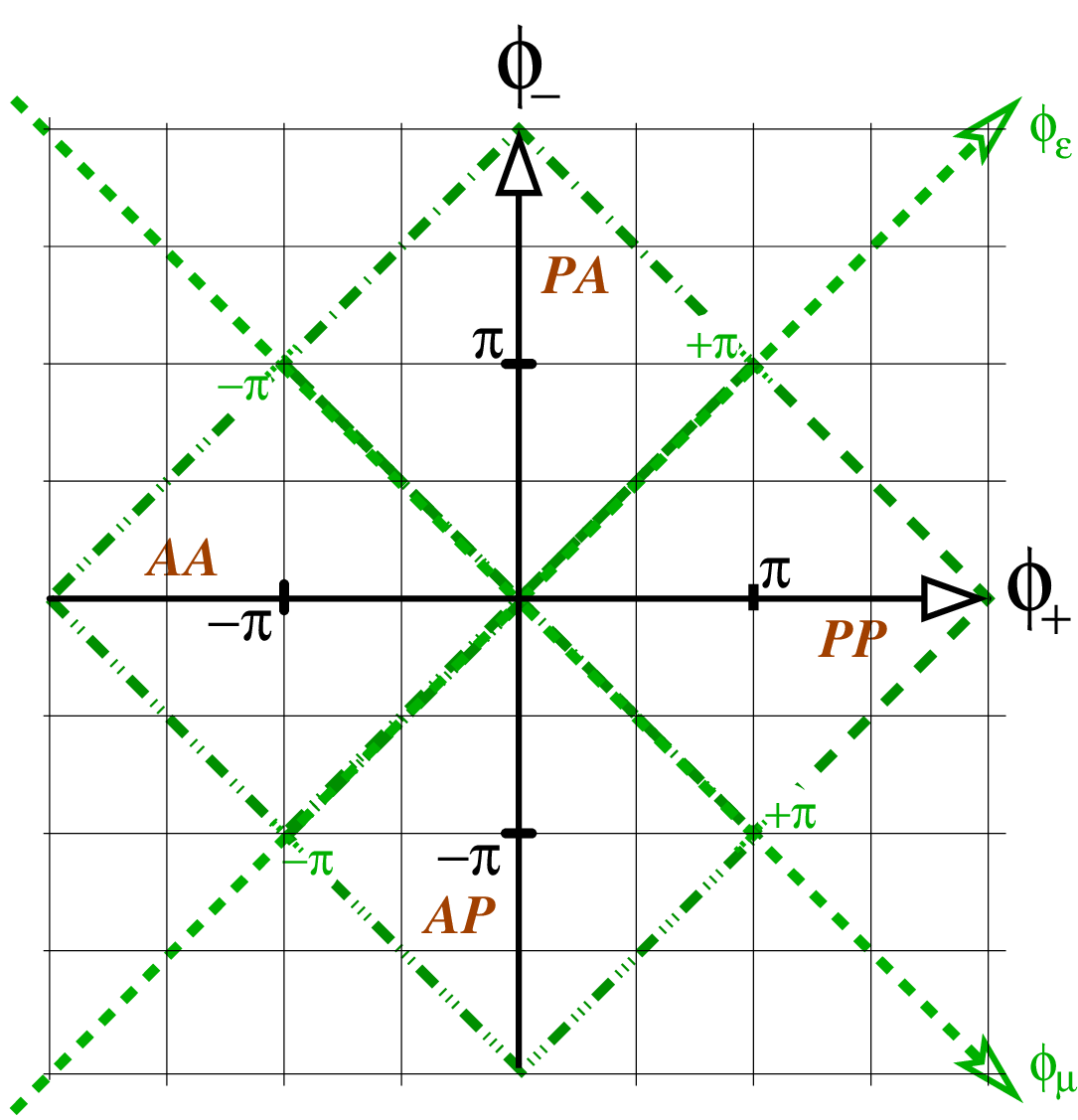}
\caption{
The complex phases ($\PhiEps$ and $\PhiMu$) of $\epsilon$ and $\mu$.
The vertical and horizontal axes (black) are sum ($\phi_+$) and 
 difference ($\phi_-$) axes; 
 the diagonal axes (green) are for $\PhiEps$ and $\PhiMu$.
The central diamond-shaped area 
 (defined simply by the ranges of $\PhiEps$ and $\PhiMu$),
 is divided into 
 doubly passive causal media (dashed box ``PP''), 
 passive-$\epsilon$ active-$\mu$ (dash-dotted ``PA''),
 doubly active (dash-dot-dotted ``AA''), 
 and 
 active-$\epsilon$ passive-$\mu$ (dash-dot-dot-dotted ``AP'').
}
\label{fig-axesetc}
\end{figure}

The properties of $\epsilon$ and $\mu$ relevant to 
 our discussion are simply their complex phases,
 i.e. $\PhiEps = \ArgPart (\epsilon)$ 
 and $\PhiMu = \ArgPart (\mu)$.
Both phases span the range $\PhiEps, \PhiMu \in \left( -\pi, \pi\right]$.
The regions where $\epsilon,\mu$ are active and passive regions
 depend on the $\PhiEps$ and $\PhiMu$ axes, 
 however 
 the NPV criteria favour the use of 
 the sums and differences,
~
\begin{eqnarray}
  \phi_+
&=&
  \PhiEps + \PhiMu
,
\\
  \phi_-
&=&
  \PhiEps - \PhiMu
,
\end{eqnarray}
and $\phi_\pm \in \left(-2\pi,2\pi\right]$.
For clarity, 
 we include both types of axes on all the figure herein; 
 but first see fig. \ref{fig-axesetc}, 
 which identifies out the different regions 
 (i.e. the doubly passive, 
 the two distinct singly active, 
 and the doubly active region).

%
\section{Comparison of Conditions for NPV}\label{S-DL12}

Depine and Lakhtakia \cite{Depine-L-2004motl} derived 
 the most general condition for NPV propagation.
This requires that $\vec{P} \cdot \vec{k}<0$, 
 with Poynting vector $\vec{P}$ and wavevector $\vec{k}$.
The definition can also be extended to embrace moving media, 
 where $\vec{P}$ is replaced by the 
 electromagnetic energy momentum tensor\cite{McCall-2008meta}. 

Taken from Depine and Lakhtakia \cite{Depine-L-2004motl}
 eqn. (12), 
 we see that NPV propagation occurs provided
 ~
\begin{eqnarray}
  \epsilon_r
  \left| \mu \right|
 +
  \mu_r
  \left| \epsilon \right|
&<&
  0
.
\label{eqn-D12i}
\end{eqnarray}
In terms of our $\phi_\pm$, 
 the condition can be written as
~
\begin{eqnarray}
  \cos \frac{\phi_+}{2}
  \cos \frac{\phi_-}{2}
&<&
  0
.\label{eqn-D12iequivalent}
\end{eqnarray}
This condition, 
 as noted by Skaar \cite{Skaar-2006ol,Skaar-2006pre}, 
 is in fact valid beyond the regime of passive media
 (although that was not so claimed in \cite{Depine-L-2004motl}). \footnote{
Subsequently \XARXIV{0812.0171} has pointed out that this condition
 fails for one counter-example involving active media
 producing a negative refractive index.
To clarify,
 our remark here is based solely on the fact that the Depine and Lakhtakia (DL)
 derivation did not restrict the signs of the real or imaginary parts
 of $\epsilon, \mu$ in any way.
The basis of the claims in arXiv:0812.0171 are as follows:
Chen et al. 2005 (CFW) propose an active medium
 with a negative refractive index, 
 which (as \XARXIV{0812.0171} points out)
 violates the DL condition.
However, 
 the CFW medium is \emph{right handed} -- 
 but DL's derivation addresses the NPV case of a \emph{left handed} medium.
Thus whether the CFW medium can be used a counter-example
 to the DL condition is debatable.}
Skaar, 
 however, 
 also noted that stability requirements mean that 
 that the product $\epsilon \mu$ 
 has no odd-order zeros in the upper half-plane.

\begin{figure}
\includegraphics[width=0.75\columnwidth]{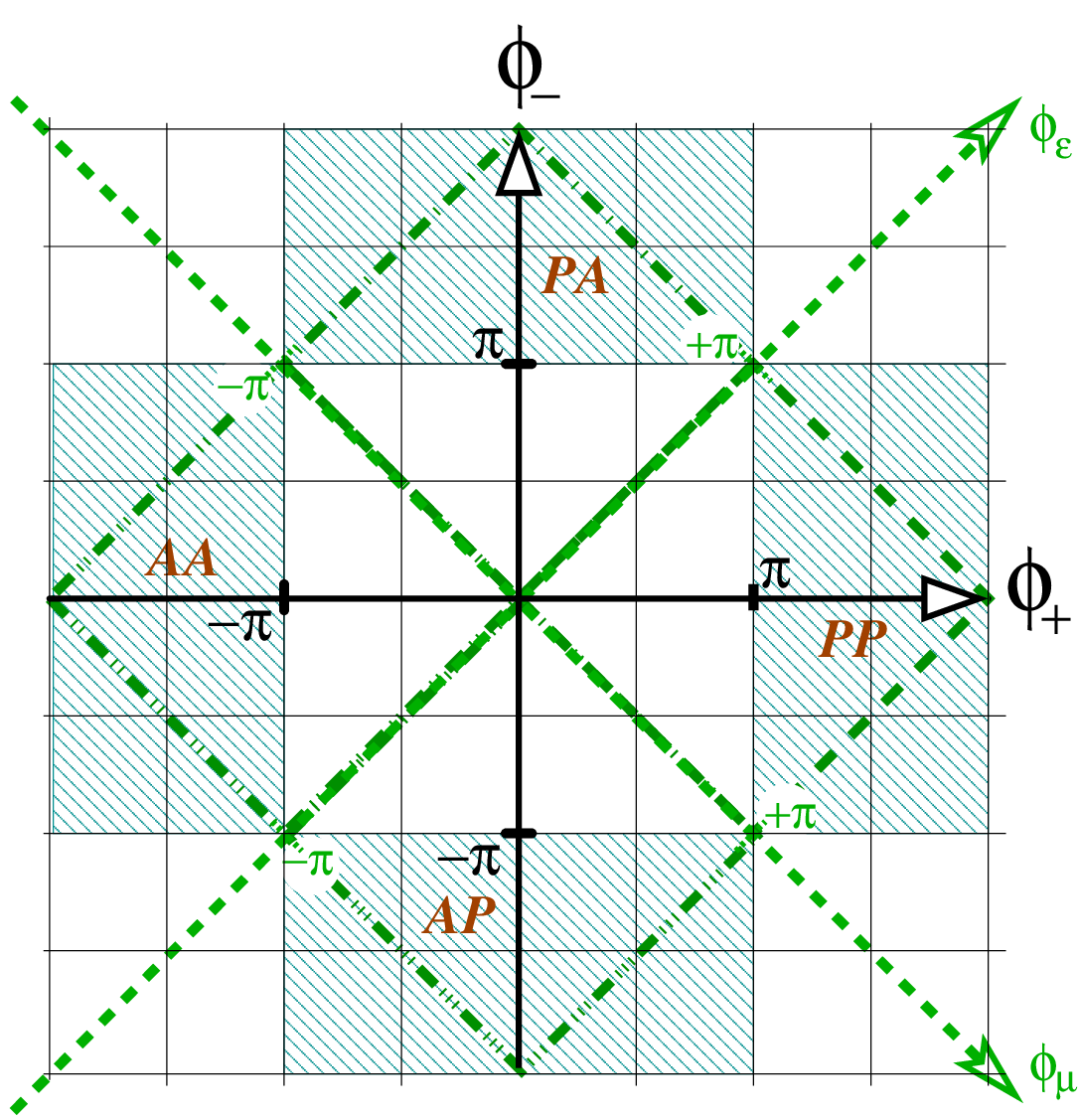}
\caption{
The region where Depine and Lakhtakia's condition 
  $\epsilon_r
  \left| \mu \right|
 +
  \mu_r
  \left| \epsilon \right|
 < 0$, 
is satisfied.
The axes and labeled regions are explained in fig. \ref{fig-axesetc}.
}
\label{fig-D12}
\end{figure}

The regions for which $\PhiEps$ and $\PhiMu$ 
 satisfy the condition is
 shown in fig. \ref{fig-D12}.
The symmetry of those regions demonstrate that it is the 
 relationship between $\PhiEps$ and $\PhiMu$ that is important, 
 rather than specific values.
However, 
 the particular phases do become more important
 if we consider the case of purely passive media.

\begin{figure}
\includegraphics[width=0.75\columnwidth]{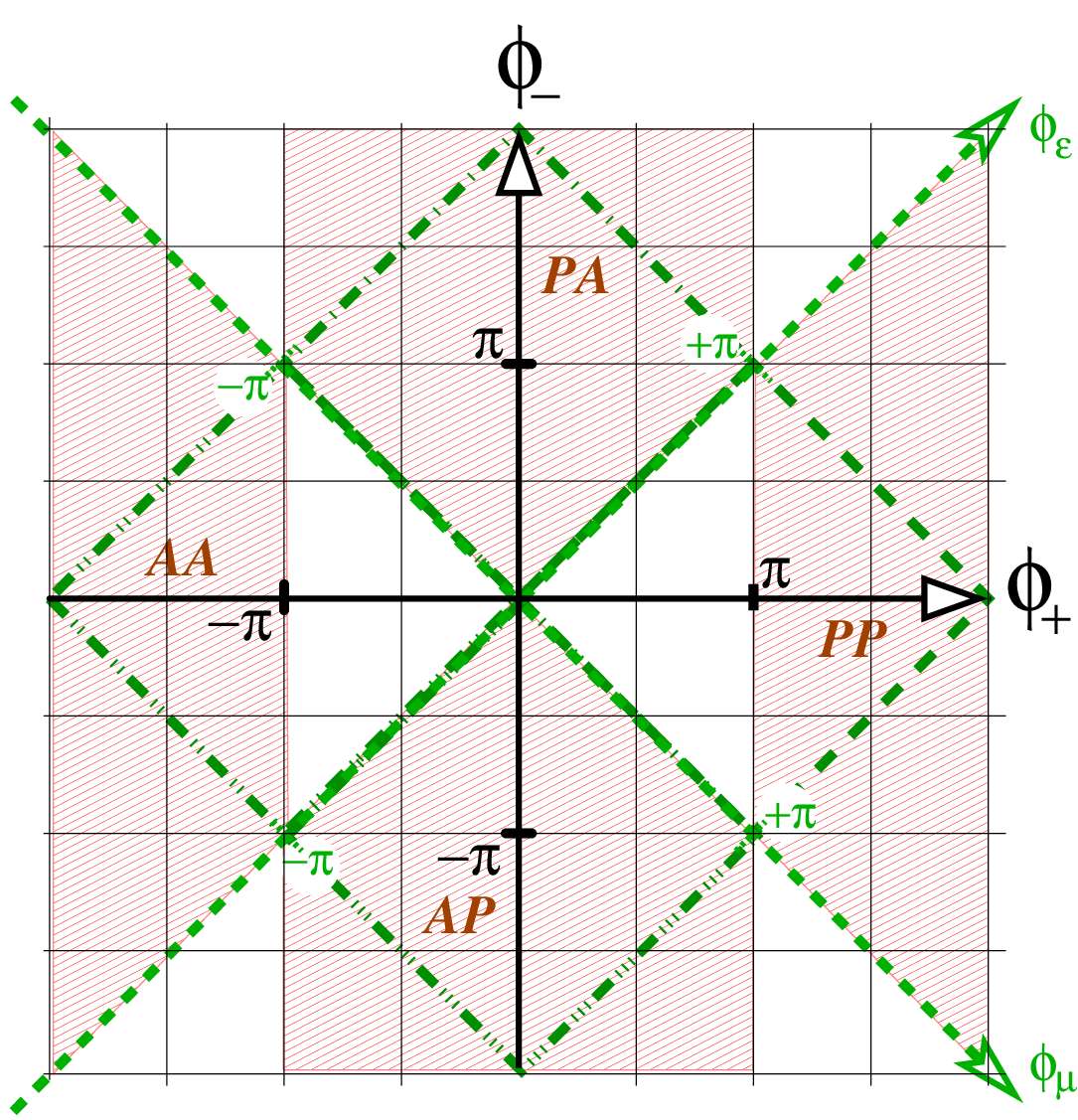}
\caption{
The region where McCall \emph{et al.}'s condition 
 $  \left[ 
    \left| \epsilon \right|
   - 
    \epsilon_r
  \right]
  \left[ 
    \left| \mu \right|
   - 
    \mu_r
  \right]
 > 
  \epsilon_i
  \mu_i$,
is satisfied.
The axes and labeled regions are explained in fig. \ref{fig-axesetc}.
}
\label{fig-D13}
\end{figure}

An alternative for NPV propagation has been given by 
 McCall \emph{et al.} \cite{McCall-LW-2002ejp}  
 (their eqn. (12)), 
 as
 ~
\begin{eqnarray}
  \left[ 
    \left| \epsilon \right|
   - 
    \epsilon_r
  \right]
  \left[ 
    \left| \mu \right|
   - 
    \mu_r
  \right]
&>&
  \epsilon_i
  \mu_i
.
\label{eqn-D13i}
\end{eqnarray}
Using $\PhiEps$ and $\PhiMu$,
 this condition becomes
\begin{eqnarray}
  \cos \frac{\PhiEps+\PhiMu}{2}
  \sin \frac{\PhiEps}{2}
  \sin \frac{\PhiMu}{2}
&<&
  0
.
\label{eqn-D13iequivalent}
\end{eqnarray}
Depine and Lakhtakia \cite{Depine-L-2004motl} 
 have showed that 
 for passive media eqn. (\ref{eqn-D12i}) is equivalent 
 to McCall \emph{et al.'s} eqn. (\ref{eqn-D13i}).
Further, 
 the equivalence also holds for for doubly active media
 (i.e. for $\epsilon_i \mu_i > 0$),
 but eqns. (\ref{eqn-D12i}) and (\ref{eqn-D13i})) are not equivalent
 in the singly active cases.
Fig. \ref{fig-D13} depicts 
 the region in which $\epsilon_i \mu_i > 0$ holds;
 of course it is only valid to the left and right hand sides of the origin;
 those being the doubly active (AA) and doubly passive regions (PP).
In those regions figs. \ref{fig-D12}
 and \ref{fig-D13} agree, 
 as expected.
The reason why the RHS of eqn. (\ref{eqn-D13i})
 is not useful for singly active media is easily seen --  
 there either $\epsilon_i$ or $\mu_i$ will be negative,
 whilst the LHS is always positive.

%

\begin{figure}
\includegraphics[width=0.75\columnwidth]{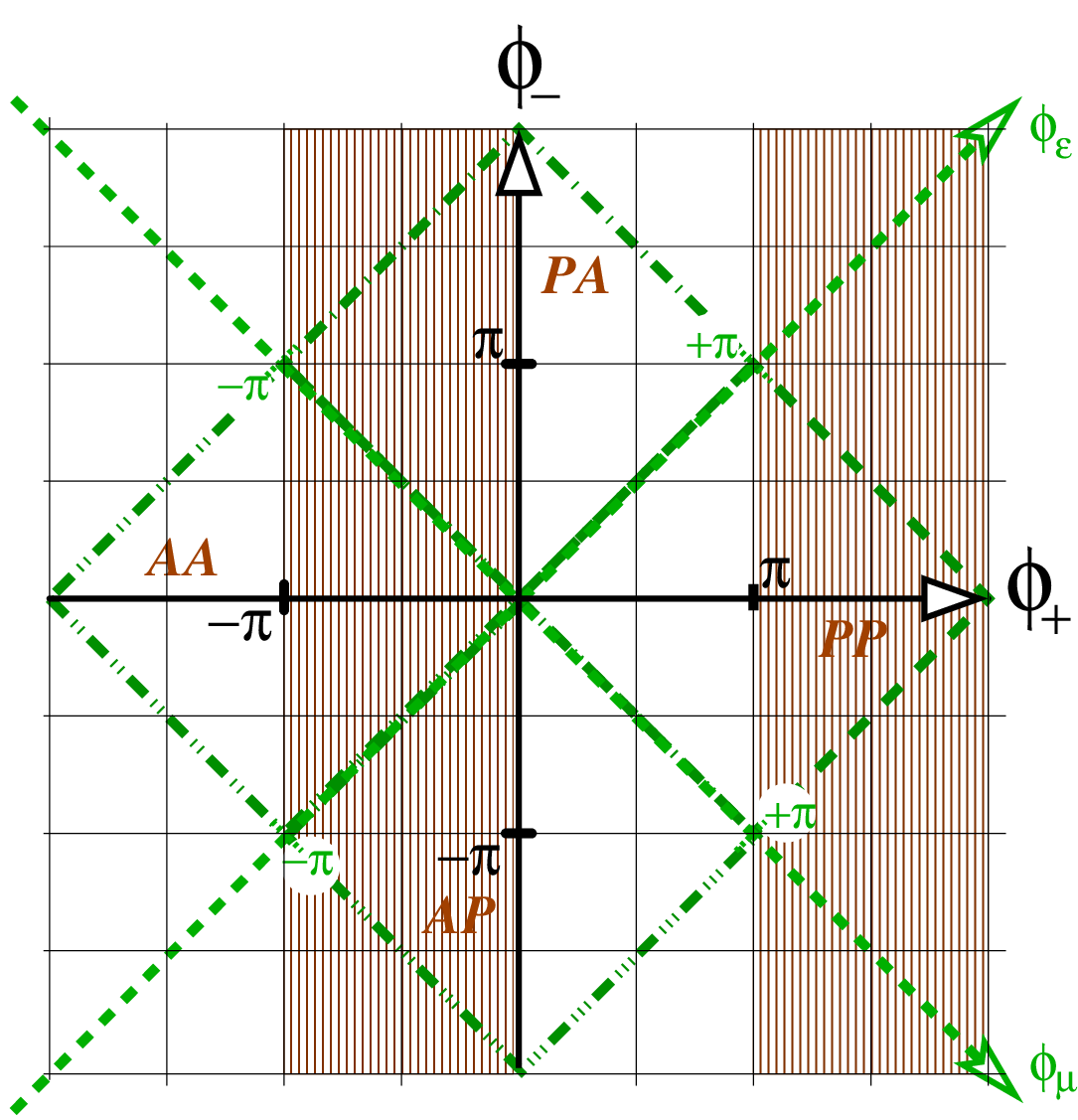}
\caption{
The region where Ruppin's condition
 $  \epsilon_r
  \mu_i
 +
  \epsilon_i
  \mu_r
 < 0$, 
is satisfied.
The axes and labeled regions are explained in fig. \ref{fig-axesetc}.
}
\label{fig-D18}
\end{figure}

Another commonly used condition is one that applies solely 
 to the imaginary part of $n^2$; 
 i.e. $\ImagPart (n^2) < 0$.
This condition has been attributed to R. Ruppin \cite{Ruppin-via-LCW}, 
 but it is worth noting that 
 Klar \emph{et al.} \cite{Klar-KDS-2006stqe}
 also provide support for it using the work of 
 Ziolowski and Heyman \cite{Ziolkowski-H-2001pre}:
Klar \emph{et al.}'s eqn. (3)
 generates the condition when 
 the real and imaginary parts of $n$ have opposite signs.
The condition is
 ~
\begin{eqnarray}
  \epsilon_r
  \mu_i
 +
  \epsilon_i
  \mu_r
&<&
  0
,
\end{eqnarray}
or equivalently
\begin{eqnarray}
  \sin \phi_+
&<&
  0
.
\label{eqn-D18i}
\end{eqnarray}
Further, 
 $\ImagPart (n^2) < 0$ was also used 
 recently by Stockman \cite{Stockman-2007prl} 
 as part of a more complicated integral-based 
 condition for negative refraction.
On fig. \ref{fig-D18} we show
 the region in which this condition holds,
 but note that it only applies to passive media\cite{Skaar-2006ol}; 
 indeed, the only agreement with fig. \ref{fig-D12}, 
 is in the passive  (PP) region.

%
\section{Conclusion}\label{S-conclusion}

We have compared the regions under which three commonly used conditions 
 for NRI are satisfied, 
 and so gained a clearer picture of their regimes of validity.
The general condition derived by
 Depine and Lakhtakia \cite{Depine-L-2004motl}
 has given us a basis on which to identify 
 the parameter ranges for which NPV propagation
 occurs for doubly (AA) and singly (PA and AP) active media.
Our graphical comparison with the other conditions 
 present in the literature emphasizes
 that they are valid only in the purely passive (PP) case.
The conditions and regimes of gain and/or loss that have been discussed
 can also be applied 
 to an acoustic model based on bulk modulius and mass density parameters.

%
\acknowledgments
The authors acknowledge financial support from the
 Engineering and Physical Sciences Research Council
 (EP/E031463/1).


%

\bibliography{/home/physics/_work/bibtex}

\end{document}